# APPLICATION OF THE WAR OF ATTRITION GAME TO THE ANALYSIS OF INTELLECTUAL PROPERTY DISPUTES

Manuel G. Chávez-Angeles, MPA/ID, Sc.D.
Universidad de la Sierra Sur.

Patricia S. Sánchez-Medina, Sc.D.
Instituto Politécnico Nacional, CIIDIR-Oaxaca.

## ABSTRACT

*In many developing countries intellectual property infringement and the commerce of pirate goods is an entrepreneurial activity. Digital piracy is very often the only media for having access to music, cinema, books and software. At the same time, bio-prospecting and infringement of indigenous knowledge rights by international consortiums is usual in places with high biodiversity. In these arenas transnational actors interact with local communities. Accusations of piracy often go both ways. This article analyzes the case of southeast Mexico. Using a war of attrition game theory model it explains different situations of intellectual property rights piracy and protection. It analyzes different levels of interaction and institutional settings from the global to the very local. The article proposes free IP zones as a solution of IP disputes. The formation of technological local clusters through Free Intellectual Property Zones (FIPZ) would allow firms to copy and share de facto public domain content for developing new products inside the FIPZ. Enforcement of intellectual property could be pursuit outside of the FIPZ. FIPZ are envisioned as a new type of a sui generis intellectual property regime.*

## 1.INTRODUCTION

During the XIII Century the *pecia* system was developed in Paris. This was a system of book production controlled by the University that flourished on the left bank of the Seine, around the cathedral of Notre Dame. The *pecia* system allowed university members to rent the texts on which the university masters lectured, in the form of unbound booklets or quires (*peciae*) so that they could be copied. The bookmen who rented out the booklet were known as 'stationers'; they were licensed by the university and had to swear corporate and individual oaths to this institution. Through the *pecia* system the university achieved some measure of control over the production and sale of academic texts, mainly the texts of the authorities on which the master lectured and the new work written by the masters themselves. In 1316 a university provision declared that any established *libraire* selling books had to swear the university oath, a regulation confirmed by royal authority, thereby establishing a university-controlled monopoly of the commercial book trade. Nevertheless, the grip of the university on the book trade was never complete. During XIV and XV centuries an increasing number of *libraires* escaped university control by setting up shop in parts of the city well outside the neighborhoods where the book trade was traditionally located. The Parisian *libraires* beyond *Ile de la Cité* and the *Quartier Latin* outside of the University control may be one of the first pirate industries in the West (Croenen, 2006).

In the XVI century, the discovery of the Americas brought new texts to the Old World. Just after the conquest of Tenochtitlan, Hernán Cortes, the Spanish Conquistador, wrote to the King of Spain Carlos V to send physicians to study herbal indigenous medicine. In 1570 Dr. Francisco Hernández, physician of King Felipe II, arrived from Spain starting a long history of exchange between European and Indigenous American medicine. From the chronicles of Franciscan frays we know that Aztecs had a well established medical professions with physicians (*tlama*), surgeons (*texoxolatitl and tesor*) and pharmacists (*papini, panamacani and panamacoyan*). In 1552, *Martín de la Cruz* wrote a volume about herbal indigenous medicine use in nowadays Mexico. Originally written in Nahuatl and later translated to Latin as *Libellus de medicinalibus indorum herbis* ("Little Book of the Medicinal Herbs of the Indians") were a codex on indigenous phramacopea. Perhaps the earlier written document of indigenous pharmacopeia is the *Codex Bodley*, now in Oxford University's Bodleian Library.

Yochai Benkler (2006) proposed the idea of three different layers embedded in communication systems that together make communication possible: (1) the physical devices and network channels necessary to communicate; (2) the existing information and cultural resources out of which new statements must be made; and (3) the logical resources the software and standards necessary to translate what human beings want to say to each other into signals that machines can process and transmit. Benkler also made the question whether there will, or will not, be a core common infrastructure that is governed as a commons and therefore available to anyone who wishes to participate in the networked information environment outside of the market-based, proprietary framework.

Evolutionary economics considers the central features of technological systems to be economic competences, clustering of resources, and institutional infrastructure. A technological system is defined as a dynamic network of agents interacting in a specific economic area under a particular institutional framework and involved in the generation, diffusion, and utilization of technology. Technological systems are defined in terms of knowledge flows rather than flows of ordinary goods and services. In the presence of creative minds and sufficient critical mass, such networks can be transformed into development blocks, synergistic clusters of firms and technologies that enhance economic growth (B. Carlsson and R. Stankiewicz, 1991).

This paper is about commerce and exchange of goods from the second layer mentioned by Benkler, and how conditions on the first and third layers affect commerce and exchange of information and cultural resources. At the same time, is about those networks of agents embedded in technological systems that interact looking for economic benefits. Specifically, the paper analyzes situations in the intellectual property regime in two arenas: (i) the digitalization of cultural industries (e.g. music, cinema); and (ii) the bioprospection of pharmaceuticals.

## 2.MASS MEDIA AND TRADITIONAL CULTURE: A STORY OF TWO CULTURES.

In Mexico the evidence of the relationship between mass media and science and technology public perception is ambiguous. Table 1 shows ordinary least square (OLS) correlations estimated with data from Mexico's National Institute of Statistics, Geography and Informatics (*Instituto Nacional de Estadistica, Geografía e Informática, INEGI*). The multivariate analysis correlates the public acceptance of patents and the public acceptance of complementary and alternative medicine (CAM).

In the United States CAM is an umbrella term given to a collection of disparate healing practices. Some of the therapies like herbs or acupuncture are well known and fairly used. Two different therapies in terms of technique and theory, herbs and acupuncture are closely related within the holistic health system of oriental medicine (Ruggie, 2004). In this article the acronym CAM is applied to a broad group of methods mentioned in the Public Perception of Science and Technology National Survey (*Encuesta sobre la Percepción Pública de la Ciencia y la Tecnología en México*), encompassing chiropractics, acupuncture, homeopathy and traditional Mexican herbal remedies (*limpias*). The information on public acceptance of patents uses information from a question asking if scientist should take more care in patenting their discoveries. Two proxies for utopian and dystopian perception of science and technology were built. The variable *utopian perception* use information from a question asking the opinion if science would ever find the cure to HIV; the variable coded as *dystopian perception* use information from a question asking the opinion if science and technology would someday destroy de planet.

| | Dependent variables | |
|---|---|---|
| **Table 1.**<br>**Ordinary Least Squares of public acceptance of patents and public acceptance of CAM** | | |
| | Public acceptance of patents | Public acceptance of CAM |
| Public acceptance of Patents | | -0.2095*<br>(0.091) |
| Public acceptance of CAM | -0.2003**<br>(0.90) | |
| Schooling | 0.0077<br>(0.019) | -0.0306<br>(0.015)* |
| Gender (women) | -0.0780<br>(0.079) | -0.0291<br>(0.065) |
| Age | 0.0002<br>(0.002) | -0.0008<br>(0.0017 |
| TV watching<br>(hours per week) | -0.0007<br>(0.003) | 0.0165**<br>(0.0033) |
| Radio listening<br>(hours per week) | 0.0473**<br>(0.005) | 0.0312<br>(0.003) |
| Newspaper reading<br>(days per week) | 0.0889**<br>(0.016) | -0.0159<br>(0.0127) |
| Magazine reading<br>(days per two weeks) | 0.0541**<br>(0.020) | 0.0312*<br>(0.0146) |
| Internet surfing<br>(hour per week) | -0.0139**<br>(0.002) | -0.0041*<br>(0.0023) |
| Computer | 1.2458**<br>(0.110) | -0.1605<br>(0.1106) |
| Utopian perception | 1.7216**<br>(0.127) | 0.6924**<br>(0.124) |
| Dystopian perception | 0.0501<br>(0.81) | 0.5972**<br>(0.065) |
| constant | -1.3282<br>(0.196)* | 0.2767<br>(0.173) |
| *Statistically significant at 90% confidence interval.<br>**Statistically significant with 95% of confidence interval. | | |

Regarding the exposure of Mexico's public opinion to mass media, Table 2 shows estimates done using data from Mexico's 2009 National Survey on the Use of Time (*Encuesta Nacional sobre el Uso del Tiempo, ENUT*). While 60.7 percent of households own a mobile phone, making mobile technology the main driver in digital inclusion; 21.1 percent of households are still in a complete media blackout or extreme digital poverty, without access to wire or mobile telephones, the Internet or cable TV.

| Table 2. Connectivity Diffusion per Households (%) | | | | |
|---|---|---|---|---|
| **Telephone** | **Mobile** | **Cable TV** | **Internet** | **Extreme Digital Poverty\*** |
| 40.3 | 60.7 | 27.7 | 18.1 | 21.1 |
| Data: 2009 National Survey on the Use of Time (ENUT), INEGI * Any of the former connections are available in the household | | | | |

Although Mexico's digital poverty, 58.9 percent of hours devoted to leisure are spend on mass media provided activities like watching TV, surfing the Internet, reading magazines or books; 28.1 percent of hours devoted for leisure are spend on community and social gathering; 6.7 percent of hours are spend practicing some sport; 4.2 percent of hours are devoted to ludic activities, either playing and instrument or a video game; and only 2.1 percent of hours are devoted to cultural activities like visiting museums, concerts or exhibitions (Table 3).

| Table 3. Demand for Leisure | | | |
|---|---|---|---|
| **Time Devoted to Different Activities (%)** | **Total Hours** | **Women** | **Men** |
| Community and social interaction (e.g. parties, church, civic events, phone calls, chat) | 28.1 | 30.4 | 25.8 |
| Culture (e.g. museums, parks, cinema, exhibitions, theater, concerts) | 2.1 | 2 | 2.2 |
| Playing non-sports (e.g. music instruments, dancing, painting, video games, board games) | 4.2 | 3.2 | 5.1 |
| Sports (e.g. soccer, basketball, swimming, boxing, karate, jogging, biking) | 6.7 | 4.9 | 8.4 |
| Mass media (e.g. books, magazines, newspapers, TV, radio, Internet) | 58.9 | 59.5 | 58.5 |
| Data: 2009 National Survey on the Use of Time (ENUT), INEGI | | | |

Taking a closer look to mass media consumption, Table 4 shows that 68.2 percent of hours are spend watching TV; 11.5 percent of hours are spend listening some kind of audio either listening the radio or music from a stream; 10.7 percent of hours are devoted to reading and only 9.6 percent of leisure time is spend in the Internet. Mexico in this regards is neither a country of readers nor of hackers; it is a country of TV watchers. The low proportion of hours devoted to reading and surfing the web could be related to illiteracy and digital poverty.

| Table 4. Mass media demand per good | | | | | | | | | |
|---|---|---|---|---|---|---|---|---|---|
| Total of hour | | Reading hours | | TV hours | | Audio hours | | Internet hours | |
| Number | % | Number | % | Number | % | Number | % | Number | % |
| 483,800.3 | 100 | 51940.7 | 10.7 | 329802.6 | 68.2 | 56030.15 | 11.5 | 46026.8 | 9.6 |
| Data: 2009 National Survey on the Use of Time (ENUT), INEGI | | | | | | | | | |

Regarding TV consumption, while 68.2 percent of hours (Table 4) devoted to mass media are spend watching TV only 27.7 percent (Table 2) of consumers have a connection to cable TV. According to the 2010 Census (INEGI, 2011) on this year 94.7 of households owned a TV. This evidence allow to infer that a high proportion of TV consumers are only watching channels broadcasted through VHF and UHF frequencies. These frequencies although legally own by the government are given on concession to the private duopoly TELEVISA-TVAZTECA. Both companies are famous for its soap opera productions, reality shows and reruns of old American TV series and cartoons. The biggest revenue for these networks comes from advertising, including government and political marketing. Any major transnational company in Mexico selling consumer goods advertise through their channels. Recently the results of the 2012 presidential election were contested on tribunals arguing overspending, vote buying and TV network's bias. Social media movements like *#yosoy132* also blamed the duopoly for information bias during the presidential elections. A reform to the Federal Telecommunications Law was passed by Congress in 2014.

# 3. WAR OF ATTRITION IN CREATIVE AND HIGH-TECHNOLOGY INDUSTRIES.

## 3.1. Digital music and video. A love story of police and pirates.

Intuition about intellectual property markets can be gained from the war of attrition game. The war of attrition was introduced in theoretical biology by Maynard Smith (1974), to explain animals fight for prey. Imagine a whale's carcass in an arctic beach. This is more meat than a single family of polar bears can eat. Nevertheless, the bear family that first arrive to (or found) the carcass will have to fight other families of polar bears that arrived to (or found) the carcass after them. If they allow for other family to approach and feed, the incumbents risk to be vanish from the carcass. Each time a new family arrives the incumbents will have to fight in order to preserve its right to feed from the whale, or will have to leave if their contestants are strong enough to become the new incumbents. How many bears families would feed from a single whale carcass is hard to tell. *The survival of the fittest,* or as Smith (1982) puts it, *Darwinian fitness*, is the rule determining how many families of bears get the chance to feed from the carcass. In a simple fashion, Figure 2 exemplifies the game faced by both troops of bears. Q being the whale's carcass (could be put as pounds or calories in a whale carcass in order to have an unit), the dominant strategy for both troops is to fight as long as *Q* is positive. Both troops would get *Q/2*.

| Figure 1. War of Attrition in the arctic beach | | | |
|---|---|---|---|
| | | Troop1 | |
| | | Fight | Leave |
| Troop 2 | Fight | Q/2, Q/2 | Q, 0 |
| | Leave | 0, Q | 0,0 |
| With any Q>0 the dominant strategy is to compete | | | |

Oligopolists are affected by many variables they cannot observe or estimate precisely: their own cost function, the cost function of their rivals, the state of demand or the potential of the market, and their rivals' strategic decisions. To the extent that some pieces of information are private we must envision market interaction as a game with asymmetric information. Riley (1980) and Kreps and Wilson (1982) introduced as asymmetric-information version to the realm of economics, focusing on predatory pricing behavior. In cultural industries, the war of attrition is taken through legal means, as well as through price wars.

In industrial organization is possible to distinguish three kinds of behavior by incumbents in the face of an entry threat (Tirole 1995):

a) Blockade entry: The incumbents compete as if there were no threat of entry. Even so, the market is not attractive enough to entrants.
b) Deterred entry: Entry cannot be blockade, but the incumbent modify their behavior to successfully thwart entry.
c) Accommodate entry: The incumbent find it (individually) more profitable to let entrants enter than to erect costly barriers to entry.

Figure 2 shows this behavior in a simple way, where $D_1$ and $D_2$ are the blockade-deterrence cost for pirates and industry respectively. Dominant strategies would depend on the value of $D_1$ and $D_2$. Under asymmetric information, $D_1(D_2)$ and $D_2(D_1)$ are the blockade-deterrence strategies for pirates and industry respectively. *Darwinian fitness* would be replaced by a combination of rationality and costs structure.

| Figure 2. War of Attrition in cultural industries | | | |
|---|---|---|---|
| | | Industry | |
| | | Blockade/Deterrence | Accommodate |
| Pirate | Blockade/Deterrence | Q/2-D₁, Q/2- D₂ | Q, 0 |
| | Accommodate | 0, Q | 0,0 |
| Dominant strategies depends on $D_1$ and $D_2$. | | | |

Now imagine an infinite population of individuals taking upon digital production (pirates and non-pirates) as an entrepreneurial activity. Let $\pi$ be the profit for digital production:

$$\pi_i = p \; [Q/n] - c - D_i \; ; \; \pi_m = pQ - c \qquad (1)$$

where $\pi_i$ are profits for each firm on competitive equilibria and $\pi_m$ is monopoly profit; $p$ is the price of the pirate product and $Q$ the total demand for that product; c is the cost of production and $n$ is the total number of producers in the industry. Figure 3 shows the normal form for two firms

| Figure 3. War of Attrition in cultural industries. Piracy as an entrepreneurial activity. | | | |
|---|---|---|---|
| | | Industry | |
| | | Blockade/Deterrence | Accommodate |
| Pirate | Blockade/Deterrence | $\pi_1$ , $\pi_2$ | $\pi_m$ , 0 |
| | Accommodate | 0, $\pi_m$ | 0,0 |
| Dominant strategies depends on $D_1$ and $D_2$. | | | |

The total number n of producers (pirates and non-pirates) would be reach where $\pi = 0$ :

$$n = pQ / D_i + c \qquad (2)$$

The higher the cost of blockade/deterrence the lower the number of producers. The cost of production is not an important variable given that it has been decreasing since the invention of the magnetic tape in the 1950s. In fact, there is not doubt that the current shift in cultural industries and the mushrooming of piracy is due to the plummeting of production costs after the invention of MP3 file format. Independent record labels can also be benefiting from the change in costs structure, mainly by lower production and distribution costs.

Now assume the existence of copyright in a dynamic game with infinite periods of time where the profit for the pirates in the black market is given by:

$$\pi_P = \Sigma_0^T \ p \ \boxed{Q}/n \ \square \qquad\qquad (3)$$

and for the industry is given by:

$$\pi_I = \pi_c + \pi_m = \Sigma_0^T \ p \ \boxed{Q}/n \ \boxed{+} \Sigma_T^\infty \ pQ \qquad (4)$$

where $\pi_c$ and $\pi_m$ are profits on a competitive market and on a monopoly respectively; T is the period where competition leaves the market.

| | | Industry | |
|---|---|---|---|
| Figure 4. War of Attrition in cultural industries. Piracy as an entrepreneurial activity. | | | |
| | | Blockade/Deterrence | Accommodate |
| Pirate | Blockade/Deterrence | $\pi_P$ - D$_P$, $\pi_I$ -D$_I$ | $\pi_P$ , 0 |
| | Accommodate | 0, $\pi_I$ | 0,0 |

Clearly $\pi_I > \pi_P$ . In this regards, how far is *T* does not matter as long as there are potential monopolistic gains ad infinitum after *T*. In the current state of affairs this potential monopolistic gains are given by copyright law. Once again only the costs $D_P$ and $D_I$ can change the dominant strategy for both (pirates and non-pirates) to enter the market.

Using equation 4 we performed a very simple lineal programming simulation. We assumed that *n* decreases constantly and that $D_I$ increases also constantly. Our results (Figure 6) suggest that (at least under this assumptions), the industry would have to bear some economic loss before obtaining monopoly profits. These losses could be so deep that firms would prefer to accommodate before obtaining monopolistic power.

**Figure 5.**
**Profit for industry without deterrence (*Di=0*) cost and constant decrease of pirate's population (n)**

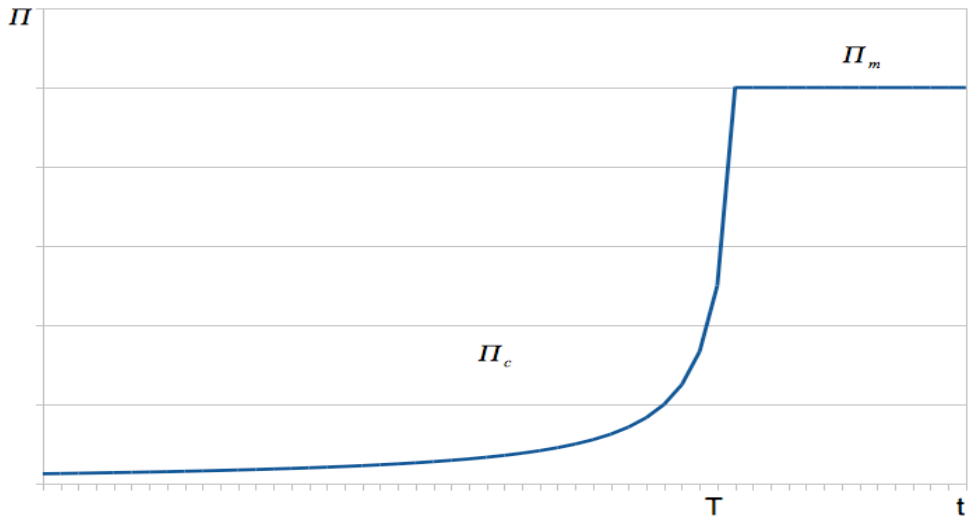

**Profit for industry with constant decrease of pirate's population and constant increase in the cost of blockade/deterrence (*Di>0; Di'>0*) .**

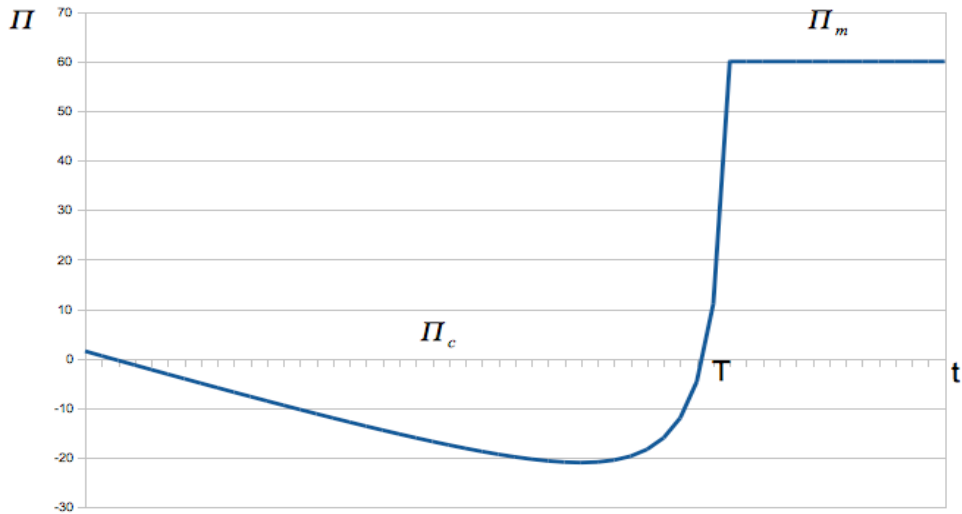

In 2005, the OECD estimated that international trade in counterfeit/pirated products amounts to USD 200 billion, excluding digital products. The Anti-Counterfeiting Trade Agreement (ACTA) was signed on the 1st October 2011. ACTA were intended to combat Intellectual Property Rights (IPRs) infringements, namely counterfeiting and piracy, by enhancing international cooperation and enforcement. The ACTA is intended to accelerate IPR enforcement, and make it more effective, so as to counter growth in counterfeiting and piracy. ACTA was negotiated between the EU and its Member States, the US, Australia, Canada, Japan, Mexico, Morocco, New Zealand, Singapore, South Korea and Switzerland. Once the agreement has entered in force, any member of the World Trade Organisation (WTO) may apply to join. The EU were joined on 26 January 2012 in Tokyo by representatives of 22 EU member states, however, the signatures need to be followed by ratification for ACTA to enter into force. (EU Parliament 2012)

| Table 5. Demand for Publishing and Recording (%) | | | | | |
|---|---|---|---|---|---|
| **Place of purchase** | **Goods** | | | | |
| | **Books** | **Newspapers** | **Magazines** | **Audio** | **Video games** |
| "Farmers" market | 1.2 | 1.4 | 3.02 | 9.5 | 5.7 |
| Street Market (*tianguis*) | 3.6 | 0.3 | 1.6 | 22.4 | 10.2 |
| Street Vendor | 8.8 | 64.9 | 26.6 | 34.2 | 7.9 |
| Convenience store | 0.2 | 7.8 | 6.7 | 0.5 | 0 |
| Specialized Stores | 65.2 | 19.2 | 42.9 | 19.1 | 38.6 |
| Supermarkets | 3.8 | 0.7 | 14.1 | 6.9 | 12.5 |
| Department stores | 6.9 | >0.1 | 2.1 | 2.4 | 13.6 |
| Bought outside the country | 0.4 | >0.1 | 0.2 | 0.2 | 3.4 |
| Club stores | 0.8 | >0.1 | 0.2 | 0.5 | 5.7 |
| Person | 8.02 | 5.3 | 1.9 | 3.9 | 2.3 |
| Internet | 0.8 | >0.1 | 0.3 | >0.1 | 0 |
| Data: 2010 National Survey of Income and Consumption (ENIGH), INEGI | | | | | |

The 4th of July 2012, 478 members of the European Parliament voted against the of Anti-

Counterfeiting Trade Agreement (ACTA), 39 in favor, and 165 abstained, meaning the agreement will not enter into force in the European Union. ACTA aimed to more effectively enforce intellectual property rights on an international level. However, opponents were concerned that ACTA would have favoured large companies' interests at the expense of citizens' rights. The European Commission referred ACTA to the European Court of Justice in May for a ruling on the agreement and asked Parliament to wait for its conclusions. Simultaneously, Parliament decided to press ahead with its own scrutiny of the agreement. Five committees came out against the agreement while the petition committee received a petition against ACTA signed by nearly three million people. (EU Parliament 2012)

The 11[th] of July 2012, Mexico's ambassador to Japan signed on to ACTA. The reasons Mexico did not sign this agreement earlier, was that the legislature had also rejected ACTA, and the executive branch had to get the Mexican Senate to ratify the agreement. Since Mexico joined the World Trade Organization on January 1[st] 1995, government's executive branch has traditionally support international initiatives aimed to more effectively enforce intellectual property rights. Media companies from developed countries worried that their business suffers great damage due to counterfeiting and piracy, have an important (although not always effective) ally in Mexico's law enforcement agencies. In support to intellectual property rights agreements Mexico's signed the Trade-Related Aspects of Intellectual Property Rights agreement (TRIPs) and Congress approved related legislation even before 1995.

The substantive legislation governing industrial property is contained in the Industrial Property Law (LPI), which has been in force since 1991. Its immediate precedent is the Law on the Promotion and Protection of Industrial Property, a body of law which was considered to be in accordance with the international standards of the time. The LPI regulates the legal concepts of patents, utility models and industrial designs, trade secrets, marks, advertising slogans and trade names, appellations of origin, and lay-out designs of integrated circuits. It also lays down the administrative procedures designed to protect and preserve industrial property rights and permitting the authorities to order the imposition of provisional measures aimed at the immediate suspension of allegedly violatory behavior.

The area of copyright is currently governed by the Federal Law on Copyright, which came into force in 1996, abrogating the Federal Law on Copyright of 1956. This body of legislation contains provisions relating to copyright, to the moral and economic rights of authors, to the rules governing the transfer of such economic rights, to copyright protection, to neighboring rights, and to limitations on copyright and neighboring rights. It establishes guidelines with respect to copyright in national symbols and expressions of popular culture, registration of rights, reservation of rights of exclusive use, collective administration of rights, the National Copyright Institute, procedure before the judicial authorities, conciliation, arbitration, copyright infringement, trade-related infringement and administrative appeal.

On the 15[th] of June 2006 Mexico's General Prosecutors (PGR), together with other branches of the executive branch and private corporations including TELEVISA, Business Software Association (BSA), the Motion Picture Export Association of America, Inc (MPEAA), signed the National Agreement Against Piracy. In the intellectual property area, Mexican law defines conduct that is subject to criminal action and punishable by monetary fines and imprisonment. In order to fulfill WTO agreements, Mexico decided to reform its legislation to ensure that conduct subject to criminal action in the intellectual property area was considered to constitute a serious offense, not only increasing the penalties, but also eliminating the privilege of release on bail. (PGR 2006).

In a digital divide environment where most of the piracy is done through hard copies (e.g. DVDs, CDs or USB memory sticks) the actions taken by PGR are old fashion raids. Between 2007 and 2010 PGR performed 7,242 raids; arrested 2,508 persons; confiscated 353, 432, 865 million copies of pirate items; and dismantled 714 laboratories (Table 7). Although this apparently spectacular numbers, there is no way to make proper evaluation of the effectiveness and impact of this measures. Total numbers for piracy activities remain unknown. At the same time, given the low cost of digital copying technology, is very probable that as soon a pirate lab is dismantled another one mushroomed filling its place. High-tech measures like Digital Rights Management (DRM) seems highly improbable to be effective. The control over the copy of content is loose by corporations once products are cracked and off-line trade and exchange of hard copies come into place to be played in devices without an Internet connection.

| | Table 7 Pirate products confiscated | | | | | | | |
|---|---|---|---|---|---|---|---|---|
| **Item** | **2007** | | **2008** | | **2009** | | **2010** | |
| | **Number** | **%** | **Number** | **%** | **Number** | **%** | **Number** | **%** |
| **DVDs** | 692,435 | 29.4 | 2,283,632 | 14.6 | 1,783,122 | 37.4 | 2,025,024 | 34.5 |
| **Video games** | 28,554 | 0.8 | 270,120 | 1.7 | 77,399 | 1.6 | 239,192 | 4.1 |
| **CDs** | 2,328,807 | 62.2 | 1,850,406 | 11.8 | 1,329,790 | 27.9 | 1,399,820 | 23.8 |
| **Clothing** | 115,946 | 3.2 | 143,281 | 0.9 | 1,328,135 | 27.8 | 1,433,271 | 24.4 |
| **Shoes (pairs)** | 48,519 | 1.4 | 6,941 | 0.04 | 236,536 | 4.9 | 737,080 | 12.6 |
| **Medicines** | 356,447 | 10.0 | 1,102,550 | 19.49 | 7,555 | 0.1 | 31,466 | 0.5 |
| **TOTAL** | 3,570,708 | 100.0 | 5,656,930 | 100 | 4,762,537 | 100 | 5,865,853 | 100 |
| Source: *5to Informe de Labores Procuraduría General de la República* (2011 General Prosecutor Annual Report) | | | | | | | | |

Notwithstanding the difficulty of doing a proper evaluation of traditional PGR's police activities, its raids allow to infer the proportion of copyright infringement. The higher proportion of pirate goods confiscated by item are hard copies of music and cinema, CDs, and DVDs, followed by medicines, clothing, shoes and video games on that order (Table 8). The highest proportion of CDs and DVDs, as well as the lowest proportion of video games confiscated is consistent with our estimates of demand from Table 5.

## 4.2. Bioprospecting: Situation analysis of Maya ICBG Project in Mexico.

Bioprospecting is a form of research and development (R&D) used by pharmaceutical or biotechnology firms to find and collect natural compounds necessary for the development of new drugs. It requires cooperation between bioprospecting firms and the community hosting the genetic resources and/or traditional knowledge. The host community provides basic or pure information on potential solution concepts, while the R&D firm supplies the practical capabilities for developing these solution concepts into marketable compounds and products. In this manner, primary biological information is generated and channeled through secondary R&D sector to become commercial products capable of addressing consumer needs (Sarr & Swanson, 2012).

To explore biodiversity in search for a medicine presents the major dilemma of where to start. In this process, indigenous knowledge represents an important window for narrowing the options and improving the probabilities of having a relevant discovery. It has been estimated that by consulting indigenous people bioprospectors can increase the success ratio in trial from 1 in 10,000 samples to 1 in only 2 (Prakash 2000). The history of drug discovery implies that traditional ethnobotanical approaches are some of most productive of the plant surveying methods. A group of researchers of the University of Uppsala found that the 86 percent of the plants used by Samoan healers display significant biological activity. Similarly researchers at the Ix Chel Tropical Research Foundation in Belize discovered that extracts of the plants that use by a traditional healer, Don Elijio Panti, gave rise to four times as many positive results in a preliminary laboratory test for activity against HIV that did specimens collected randomly (Cox and Balick, 1994).

Adapting the war of attrition model to patented industries like pharmaceuticals is not difficult. Particularly in cases like patent infringement on the production of HIV drugs in India the adaptation is straightforward. Nevertheless, when we consider bioprospecting from traditional knowledge in Southern Mexico (or any other part of the world with a rich biochemical public domain) to model competition is a little trickier.

| Figure 6. War of Attrition in cultural industries | | | |
|---|---|---|---|
| | | Industry | |
| | | Blockade/Deterrence | Accommodate |
| Pirate | Blockade/Deterrence | Q/2-D$_1$, Q/2- D$_2$ | Q, 0 |
| | Accommodate | 0, Q | 0,0 |
| Dominant strategies depend on $D_1$ and $D_2$. | | | |

In terms of traditional/indigenous knowledge we can imagine a population n of persons (or healers) holding some kind of knowledge from the public domain. Let $\pi_H$ be the benefit for using that knowledge:

$$\pi_H = p(Q/n) - c \qquad (5)$$

Now, consider a firm acquiring the patent from some portion of that knowledge. Let $\pi_m$ be the benefit from the monopolistic exploitation of the patent.

$$\pi_m = pQ - c - INV \qquad (6)$$

In both cases (5) and (6), *p* is the price of the medicine and *Q* the total demand for that product; c is the cost of production; INV is the cost of the investment in synthesizing the active substance and the legal cost of patenting. Figure 7 shows the normal form for the protection of public domain:

| | | Bioprospector firm | |
|---|---|---|---|
| | | Patent | Not to patent |
| Healers | Blockade/Deterrence | $\pi_H$, $\pi_m - f$ | $\pi_H$, 0 |
| | Accommodate | 0 , $\pi_m$ | $\pi_H$,0 |

Figure 7. War of Attrition in bioprospecting. Public domain vs. patent protection.

The dominant strategy is to patent and accommodate.

$f$ is the cost of entrance. In this scenario the dominant strategies is to patent and blockade as long as $\pi_H - f > 0$. Very often is observed that the dominant strategies are blockade and not to patent. On that case $f$ was so big that $\pi_H - f < 0$.

## 5.CONCLUSIONS AND POLICY PROPOSAL.

Digital technologies offer unprecedented possibilities for human creativity, global communication, and access to information. Yet in the last decade, mass media companies have developed methods of control that undermine the public's traditional rights to use, share, and reproduce information and ideas. These technologies, combined with an oligopolistic structure in the media industry and new laws that increase its control over intellectual products, threaten to undermine creativity, privacy ad free speech.

In June 2000 Australia posed a question to the Permanent Mission of Mexico on the Council for TRIPs at WTO. The question was if there are any specific exceptions to copyright under the law of Mexico to allow use of copyright material by third parties for permitted purposes (such as research, education, fair use or fair dealing?)  Are there any significant judicial decisions with bearing on this issue?  Are there any specific rules or findings concerning exceptions or limitations to copyright protection of computer software? (WTO 2001)

The mission answer that under the Federal Law on Copyright (LFDA), Title VI, merely establishes the limitations to copyright and neighboring rights. Until today this article haven't been applied:

Chapter I.  Limitation in the Public Interest

Article 147. The publication or translation of literary or artistic works shall be considered in the public interest where they are necessary for the advancement of science and national culture and education. Where it is not possible to obtain the consent of the owner of the corresponding economic rights, the Federal Executive may, through the Ministry of Public Education and either ex officio or at the request of a party, license the publication or translation in question against payment of compensatory remuneration.  The foregoing shall be without prejudice to any international treaties on copyright and neighbouring rights signed and ratified by Mexico.

In developed countries, libraries, civic organizations, and scholars have begun to turn the idea of information commons, with a wide variety of open democratic information resources now operating or in the planning stages. These include software commons, licensing commons, open access scholarly journals, digital repositories, institutional commons, and subject matter commons in areas like music, cinema and books. The basic characteristics in this commons is that they are collaborative and interactive, taking advantage of the networked environment to build information communities. Benefiting from network externalities, meaning that the greater the participation, the more valuable the resource. Many are free or low cost. Their governance is shared, with rules and norms that are defined and accepted by their constituents. They encourage and advance free expression. (Kranich, 2004)

In terms of how to use combine information on biodiversity and its pharmaceutical uses for fostering human development, an important challenge is how to create a technological system that includes high tech actors (e.g. biotechnology networks), local communities and indigenous people related with biodiversity; and that allows the free flow of information, knowledge and technology on an equitable and efficient manner. Tenenbaum and Wilbanks (2008) call such system Health Commons: "a coalition of parties interested in changing that way basic science is translated into the understanding and improvement of human health. Coalition members agree to share data, knowledge and service under standardized terms and conditions by committing to a set of common technologies, digital information standards, research materials, contracts, workflows and software. These commitments ensure that knowledge, data, materials and tools can move seamlessly from partner to partner across the entire drug discovery chain. They enable participants to offer standardized service, ranging from simple molecular assays to complex drug synthesis solutions, that others can discover in directories and integrate into their own processes to expedite development, or assemble like LEGO blocks to create new services."

A collaborative cluster would bring common solutions to common problems, offers critical mass for customization of interventions, economies of scale in operation, better access to technology and information, greater access to customer's channels and cheaper access to inputs and raw materials. The seed money would be used for to establish an organization that promotes de formation of public-private partnerships (PPP) for building infrastructure and propel R&D.

Appropriate financial institutional infrastructure is also important in fostering business development and technological innovation. The record of financial institutions in this field has been generally poor in this part of the country. Capital markets, such as venture capital, credit cooperatives and microfinance institutions play a critical role in creating SMEs. Other than arranging funding, venture capitalists should help groom business start-ups into competitive and profitable firms.

A Free Intellectual Property Zones (FIPZ) would allow firms to copy and share *de facto* public domain content for developing new products inside the FIPZ. Enforcement of intellectual property could be pursuit outside of the FIPZ. In a sense FIPZ would be a *sui generis* intellectual property regime. Manufacturing within the FIPZ would also be extent of other intellectual property regimes. On drug development, the clusters could takes advantage of the rich biodiversity and propel the geographical agglomeration of small and medium enterprises (SME) as long as the manufacturing is done within the FIPZ.

# BIBLIOGRAPHY.